\documentstyle[twocolumn,aps]{revtex}

\input psfig.sty
\begin{document}
\draft

\twocolumn[\hsize\textwidth\columnwidth\hsize\csname
@twocolumnfalse\endcsname

]

\noindent {\bf Aji and Goldenfeld Reply:} In a recent letter
\cite{aji}, we analyzed the critical dynamics of the superconducting to
normal phase transition in zero magnetic field. We explained Monte
Carlo (MC) simulation results\cite{mc} in both strong and weak
screening limits, where the dynamic critical exponent was found to be
$z_{MC} \sim 2.7$ and $z_{MC} \sim 1.5$ respectively. These results,
taken at face value were surprising, departing strongly from scaling
expectations based on model A dynamics that $z\sim 2$\cite{ffh}.  We
showed that the simulations do not measure the true dynamic exponent
$z$ and that in both the short and long range limits, provided the
screening length is smaller than the system size, the dynamic exponent
correctly inferred from the simulations is $z\sim 2$, thus removing the
discrepancy.

In the preceding Comment\cite{lid}, the author asserts that in our
theory, the vorticity has the wrong scaling dimension, inconsistent
with standard scaling theory and not supported by MC data. Second, he
asserts, without justification, that the identification of MC time with
real time is correct, and that our proposal that they are not the same
is an influence of the discreteness of the lattice.

In fact, the non-equivalence of MC time and real time is a genuine
effect, but arises due to an implementation of a dynamic MC algorithm
that does not properly account for the scaling of $J$.  Here, we show
that our theory predicts a scale dependent coupling constant $J(L)$,
and that the standard finite size scaling results are indeed satisfied,
contrary to the statement in the Comment.

In terms of the vorticity, $\vec{n} = \vec{\nabla} \times \vec{\nabla}
\theta$, the free energy is,
$
F = \beta \sum_{i,j}\vec{n_{i}}.\vec{n_{j}}G_{ij}[\lambda_{0}],
$
where the lattice Green function is
\begin{equation}
G_{ij}[\lambda_{0}] = J{(2\pi)^{2} \over
L^{3}}\sum_{\vec{k}}{\exp[i\vec{k} \cdot (\vec{r_{i}} - \vec{r_{j}})]
\over {2\sum_{m}^{3}[1 - \cos(k_{m})] + \lambda_{0}^{-2}}}
\end{equation}
\noindent Here $J$ is the coupling constant, $\lambda_{0}$ is the
screening length and $\beta = 1/k_{B}T$. In the long range case,
$\lambda_{0} \rightarrow \infty$, the $G_{ij} \sim J/|\vec{r_{i}} -
\vec{r_{j}}|$, while in the short range case, $G_{ij} \sim
J\delta(\vec{r_{i}} - \vec{r_{j}})$. The free energy density scales as
$L^{-d}$ ($F \sim (L/\xi)^{d}$), $\theta$, the phase, has a trivial
scaling dimension of $0$ and $n \sim \xi ^{-2}$. In the long range
case, since $G_{ij} \sim J/\xi$, we get $J \sim \xi^{2}/L^{3}$ where
$\xi$ is the correlation length ($(L/\xi)^{d} \sim
L^{2d}\xi^{-4}J/\xi$). In the short range case, where $G_{ij} \sim
J/L^{d}$, a similar analysis yields $J \sim \xi$. In the dynamic MC
simulations, the scaling of $J$ is not accounted for and leads to
scaling dimensions of $n$ other than $-2$. If one ignored the scaling
of $J$, the above analysis would yield  $n \sim  \xi ^{-x}$ and $\nabla
\theta \sim \xi^{1-x}$, where $x$ is $5/2$ ($(L/\xi)^{d} \sim
L^{2d}\xi^{-2x}/\xi$) in the weak screening limit and $x=3/2$ in the
strong screening limit. The superfluid density is obtained from,
$
F = \beta\int d\vec{r}\rho |\vec{\nabla}\theta|^{2}.
$
Given the scaling dimension of $\nabla \theta$ above, and assuming that
$J$ does not scale, $\rho \sim J\xi^{2x-2-d}$ ($(L/\xi)^{d} \sim \rho
L^{d}\xi^{2-2x}$), which would disagree with standard scaling theory
\cite{NG}.

This discrepancy in the scaling dimensions arises only in MC
simulations where a time dependent variable is measured, such as a
two-time correlation function. The superfluid density in static MC
simulations, for example, is obtained from the equal time
current-current correlation function in the phase representation of the
Villain model. Unlike the vorticity, the phase variable does not scale
with system size ($\theta \sim L^{0}$). This computation of the
correlation function reproduces the standard scaling form for $\rho$.

The non-equivalence of real time and MC time arises from using the same
value of $J$ for all lattice sizes $L$ \cite{meak}, and as explained in our letter,
is a consequence of the fact that in a single MC time step, regardless
of $L$, an equal change must be made to the voltage pulse produced by
the vortex loops and not the vortex loops themselves. In dynamic MC
simulations, if we scaled $J$ appropriately, we would obtain $z \sim
2$, for strongly screened interactions. However, one usually does not
know ahead of time what the scale dependence of $J(L)$ is.

In our theory, the scaling form for $\rho$ is calculated from $\rho
\sim J(L)L^{2x-d-2}$. In the case of short ranged interaction and the
physically relevant case of weak screening ($\lambda_{0} < L$), $x =
3/2$ and $J \sim L$ ($\xi \sim L$ at $T_{c}$), we get $\rho \sim
L^{2-d}$ as expected. In the long range case, where $\lambda_{0}$ is
set to infinity, the same argument yields $\rho \sim L^{2-d}$ since $x
= 5/2$ and $J \sim L^{-1}$ in this case. For the magnetic permeability,
including the scaling of $J$ implies, $\mu \sim J^{-1}L^{d-2x}$. With
this correction, we obtain the correct scaling form, $\mu \sim
L^{d-4}$, in both the strong and weak screening limit.

In summary, our analysis is fully consistent with standard results, and
explains the surprising results of the MC simulations.

We acknowledge National Science Foundation grant NSF-DMR-99-70690.

\vspace{0.1in}
\noindent Vivek Aji and Nigel Goldenfeld \\
{\it University of Illinois at Urbana-Champaign \\
Department of Physics \\
1110 West Green Street \\
Urbana, Illinois 61801-3080}

\end{document}